\documentclass[10pt]{article}
\usepackage{hyperref}
 \usepackage{footnote}
\makesavenoteenv{minipage}
\renewcommand{\title}[1]{%
    \bigskip%
    \begin{center}%
    \Large\bf #1%
    \end{center}%
    \vskip .2in}

\renewcommand{\author}[1]{%
    {\begin{center}
    #1
    \end{center}}}
\newcommand{\address}[1]{\vspace{-1.7em}\vspace{0pt}
    {\begin{center}
    \it #1 
    \end{center}}}

\begin{document}

\title{  Generalized BRST symmetry for arbitrary spin conformal field theory}

 \author{ Sudhaker Upadhyay$^a$ 
 \footnote{e-mail address: sudhakerupadhyay@gmail.com},
  Bhabani Prasad Mandal$^b$\footnote{e-mail address: bhabani.mandal@gmail.com}}

\address{  $^a$Department of Physics, Indian Institute of Technology Kanpur, Kanpur 208016, India  }
\address{$^b$Department of Physics, 
Banaras Hindu University, 
Varanasi 221005, India.  }

\begin{abstract}
We develop the finite field-dependent BRST (FFBRST) transformation for arbitrary spin-s
conformal field theories. We discuss the novel features of the FFBRST transformation
in these systems.
 To illustrate the results we consider the spin-1 and spin-2 conformal field theories in two
 examples. Within the formalism we found that FFBRST transformation connects the
 generating functionals of spin-1 and spin-2 conformal field theories in linear and non-linear gauges.
 Further, the conformal field theories
 in the framework of FFBRST transformation are also analysed in Batalin-Vilkovisky (BV) formulation
 to establish the results.
 \end{abstract}

\section{Introduction}
Conformal field theories (CFT) \cite{mald}  have been at the centre of much attention during the last seventeen years
mainly because of they provide models for genuinely interacting quantum
field theories, they describe two-dimensional critical phenomena, and they play a central
role in string theory, at present the most promising candidate for a unifying theory of
all forces.  Much attention has been given to conformal field theories in higher dimensions
due to their role in the AdS/CFT correspondence \cite{os,er}.   
  AdS$_3$/CFT$_2$ is one
of the most hot topics nowadays as it may be amenable to the integrability approach
that proved very successful especially in the case of AdS$_5$/CFT$_4$ \cite{ad}.
The AdS/CFT correspondence has also been investigated  for
scalar fields \cite{i,j,k}, gauge fields \cite{k}, spinors \cite{l}, classical gravity \cite{m} and type IIB
string theory \cite{n,o}.  
The AdS/CFT correspondence is used to calculate CFT correlators
from the classical AdS theories of vector and Dirac fields   and the connection between the AdS and boundary fields is
properly treated via a Dirichlet boundary value problem   \cite{j}.

Recently, in the framework of gauge invariant approach involving Stueckelberg fields  the   totally 
symmetric arbitrary spin-s anomalous
conformal current and shadow field are studied and gauge invariant two-point vertex 
of the arbitrary spin anomalous shadow field is also obtained \cite{mes0}. In Stueckelberg gauge frame, the two-point
gauge invariant vertex becomes the standard two-point vertex of CFT.
The logarithmic divergence of the BRST invariant action of
arbitrary spin-s canonical shadow field turns out to be BRST invariant action of arbitrary spin-s
conformal field \cite{mes}. The BRST invariant action
of conformal field interprets  geometrically the boundary values of massless AdS fields \cite{mes}.
The study of BRST quantization which helps in proving the renormalizibility of gauge theories
is extremely important in the context of CFT.

Although BRST symmetry has been discussed for conformal field theory  \cite{mes} 
 the generalization of  it  by making the parameter field-dependent, so-called FFBRST transformation, has not yet been investigated.
 The FFBRST formulation, which was introduced for the first time by Joglekar and Mandal
 \cite{bm},
 has been studied considerably in various context \cite{sdj1}-\cite{sud5}.
  For example, such formulation helps in calculating  a correct prescription for   poles in the gauge field 
  propagators in noncovariant gauges   by connecting the covariant gauges and noncovariant gauges of the 
  theory \cite{sdj1,jog}.
The celebrated Gribov problem \cite{gri,zwan} of QCD  has also been addressed  through FFBRST transformation  
in Euclidean space \cite{sb}. Further, such formulation has  been 
investigated for YM theory explaining low-energy dynamics via Cho--Faddeev--Niemi (CFN) decomposition.
 So, it is worth analysing such formulation at both classical and quantum level for conformal field theories. 
This provides a motivation for the analysis of FFBRST transformation in conformal field theory in present 
investigation.

We further like to extend our FFBRST formulation for CFT in the framework of Batalin-Vilkovisky (BV) formalism \cite{henna} -\cite{bv3} which
is one of the most powerful techniques to study gauge field theories and allows us to deal with very general 
gauge 
theories, including those with open or reducible gauge symmetry algebras.
The BV method provides a convenient way of analysing the possible violations of 
symmetries by quantum effects \cite{wein}.  It is usually used  to perform the gauge-fixing
in quantum field theory, but was also applied to other problems like analysing
possible deformations of the action and anomalies.
 The BRST-BV approach is a successful for studying the 
manifestly Lorentz invariant formulation of string theory \cite{sei}.

In this paper we generalize the FFBRST transformation for arbitrary spin-s conformal field theory
by making the parameter finite and field dependent. Within the formulation, we find that 
the   functional measure leads to a non-trivial Jacobian. This Jacobian can be exponentiated 
if this satisfies certain condition. As a result the effective action gets modified.
We compute the Jacobians for 
spin-1 and spin-2 conformal fields for particular choices  of finite field-dependent parameters.
We render that these calculated Jacobians play an important role in mapping of
linear and non-linear gauges.
The analysed BV formulation validate the results
at quantum level. For BV formulation we extend the configuration space
by introducing antifield  corresponding to each field with opposite statistics.
With such introduction of antifield the consequent 
extended action satisfies the mathematically rich quantum master equation.

The paper is presented in following manner. In section 2, we generalize the BRST transformation 
for arbitrary spin-s conformal field theory. We illustrate this generalization by 
two examples of spin-1 and spin-2 conformal fields in section 3.
We extend this formulation in the BV framework in section 4.
At the end we summarise the results.

  \section{Constructing FFBRST transformation for arbitrary spin-s conformal field theory} 
  
In this section we construct the FFBRST transformation for conformal field  theory following the method advocated in \cite{bm}.  
Let us begin with the effective action for arbitrary spin-s conformal field theory defined by\footnote{
We use the following conventions: $x^a$ denotes the coordinates in
$d$-dimensional flat space-time, while $\partial_a$ denotes the derivatives
with respect to $x^a$. Vector indices ( $a, b, c, e...$) of the Lorentz
algebra $so(d-1, 1)$ take the values $0, 1, . . . , d-1$.
We use the flat metric tensor $\eta^{ab}$  in scalar products as follows: 
$X^aY^a = \eta_{ab}X^aY^b$.} \cite{mes},
\begin{eqnarray}
S_{tot}= \int d^d x\left[\sum_{s'=0}^s {\cal L}^{s'} +\sum_{s'=0}^{s-1} {\cal L}_{FP}^{s'}  \right],
\label{tot}
\end{eqnarray}
where 
\begin{eqnarray}
{\cal L}^{s'}&=&\frac{1}{2s'!}\left(\phi^{a_1...a_{s'}}(\partial^l\partial^l)^{\nu_{s'}}\phi^{a_1...a_{s'}} -\frac{s'(s'-1)}{4}  \phi^{aaa_3...a_{s'}}(\partial^l\partial^l)^{\nu_{s'}}\phi^{bba_3...a_{s'}}  \right),\nonumber\\
{\cal L}_{FP}^{s'}&=&\frac{1}{s'!}\bar c^{a_1...a_{s'}}(\partial^l\partial^l)^{\nu_{s'}+1}c^{a_1...a_{s'}},\ \ \ \nu_{s'}=s'+\frac{d-4}{2}.
\end{eqnarray}
This effective action is invariant  under the 
usual BRST transformation for the collective fields $\varphi^{a_1a_2.....a_{s'}} (\equiv \phi^{a_1a_2.....a_{s'}}, c^{a_1a_2.....a_{s'}}, \bar c^{a_1a_2.....a_{s'}})$ for the conformal field theory compactly as follows \cite{mes}
\begin{equation}
\delta_b  \varphi^{a_1a_2.....a_{s'}} =s_b\varphi^{a_1a_2.....a_{s'}}\ \delta\lambda ={\cal R}[\varphi^{a_1a_2.....a_{s'}}] \delta\lambda,
\end{equation}
  where ${\cal R}[\varphi^{a_1a_2.....a_{s'}}]=s_b \varphi^{a_1a_2.....a_{s'}}$ is the generic Slavnov variation of the fields $\varphi^{a_1a_2.....a_{s'}}$ written collectively
  and $\delta\lambda$ is the infinitesimal anticommuting global parameter of transformation. 
  
  Now we make the parameter $\delta\lambda$ finite and field-dependent by interpolating a continuous parameter $\kappa$ through fields which is bounded between $0$ and $1$. 
The  infinitesimal field-dependent BRST transformation is constructed as follows \cite{bm}
\begin{equation}
\frac{d\varphi^{a_1a_2.....a_{s'}}(x,\kappa)}{d\kappa}={\cal R} [\varphi^{a_1a_2.....a_{s'}} (x,\kappa ) ]\Theta^\prime [\varphi^{a_1a_2.....a_{s'}} (x,\kappa ) ],
\label{diff}
\end{equation}
where the $\Theta^\prime [\varphi^{a_1a_2.....a_{s'}} (x,\kappa ) ]$ is an infinitesimal but the field-dependent parameter.
The FFBRST transformation (denoted by $\delta_f$) then can be 
obtained by integrating the above transformation from $\kappa =0$ to $\kappa= 1$, as follows:
 \begin{eqnarray}
\delta_f \varphi^{a_1a_2.....a_{s'}}(x)&\equiv & \varphi^{a_1a_2.....a_{s'}} (x,\kappa =1)-\varphi^{a_1a_2.....a_{s'}}(x,\kappa=0)\nonumber\\&=&{\cal R}[\varphi^{a_1a_2.....a_{s'}}(x) ]\Theta[\varphi^{a_1a_2.....a_{s'}}(x) ],
\end{eqnarray}
where \begin{equation}
\Theta [\varphi^{a_1a_2.....a_{s'}}(x)] = \Theta ^\prime [\varphi^{a_1a_2.....a_{s'}}(x)] \frac{ \exp f[\varphi^{a_1a_2.....a_{s'}}(x)]
-1}{f[\varphi^{a_1a_2.....a_{s'}}(x)]},
\label{80}
\end{equation}
 is the finite field-dependent parameter and $f[\phi]$ is given 
 by 
 \begin{eqnarray}
 f[\varphi^{a_1a_2.....a_{s'}}(x)]= \sum_i \int d^4x \frac{ \delta \Theta ^\prime [\varphi^{a_1a_2.....a_{s'}}(x)]}{\delta
\varphi_i^{a_1a_2.....a_{s'}}(x)} s_b \varphi_i^{a_1a_2.....a_{s'}}(x).
 \end{eqnarray} 
 This  FFBRST transformation leaves effective action of a conformal field theories
 invariant. However, the functional
 measure changes non-trivially under such finite transformation.

Now we compute the Jacobian of the path integral measure defined generically by $({\cal D}\varphi^{a_1a_2.....a_{s'}})$ 
 for an arbitrary finite field-dependent parameter, $\Theta[\varphi^{a_1a_2.....a_{s'}}(x)]$, as follows
\begin{eqnarray}
{\cal D}\varphi^{\prime a_1a_2.....a_{s'}}&=&J(
\kappa) {\cal D}\varphi^{a_1a_2.....a_{s'}}(\kappa).\label{jacob}
\end{eqnarray}
The Jacobian $J(\kappa )$ of the path integral measure is thus obtained as a functional of fields. So
we exponentiate it by defining a local functional $ S_1[\varphi^{a_1a_2.....a_{s'}} ]$ in following manner:   
\begin{equation}
J(\kappa )\longmapsto e^{{iS_1 [\varphi^{a_1a_2.....a_{s'}}(x,\kappa) ]}}.\label{js}
\end{equation}
 Preserving the quantitative (physical) changes of the functional integral in conformal field theory
 leads to the following condition \cite{bm}
 \begin{eqnarray}
 \int {\cal{D}}\varphi^{a_1a_2.....a_{s'}} (x) \;  \left [\frac{d}{d\kappa}\ln J(\kappa)-i\frac
{dS_1[\varphi^{a_1a_2.....a_{s'}} (x,\kappa )]}{d\kappa}\right ] \exp{[i(S_{tot}+S_1)]}=0. \label{mcond}
\end{eqnarray} 
The local functional $ S_1[\varphi^{a_1a_2.....a_{s'}} ]$ satisfies the following 
initial boundary condition   $S_1[\varphi^{a_1a_2.....a_{s'}} ]_{\kappa=0}=0$
to ensure $J=1$, when fields do not change.
 
The infinitesimal change in Jacobian, $J(\kappa)$, given in (\ref{mcond}) has the
explicit expression in terms of $\Theta'$ as follows 
\begin{equation}
 \frac{d}{d\kappa}\ln J(\kappa)=-\int  d^dy\left [\pm\sum {\cal R}[\varphi^{a_1a_2.....a_{s'}}(y )]\frac{
\partial\Theta^\prime [\varphi^{a_1a_2.....a_{s'}} (y,\kappa )]}{\partial\varphi^{a_1a_2.....a_{s'}} (y,\kappa )}\right],\label{jac}
\end{equation}
where, for bosonic fields,  $+$ sign is used and 
$-$   for fermionic fields.
      
Therefore, performing FFBRST transformation 
changes the exponential action of the generating functional given in conformal field theory
as following:
\begin{eqnarray}
\int {\cal D}\varphi^{a_1a_2.....a_{s'}} e^{iS_{tot}}\longrightarrow\int {\cal D}\varphi^{a_1a_2.....a_{s'}}
 e^{i(S_{tot}+S_1)},\label{ext}
\end{eqnarray}
where $S_{tot}$ is the most general effective
action for CFT given in (\ref{tot}).
To illustrate these results we would like to 
consider specific examples in the next sections.
 \section{BRST invariant  conformal fields}
  In this section, we consider the two examples of BRST symmetric conformal field theory.
  We study the construction and implementation of FFBRST transformation on these theories
  explicitly.  
  \subsection{Spin-1 conformal field}
  The BRST invariant action for spin-1 conformal field (a particular form of (\ref{tot})) in linear gauge is given by
  \begin{eqnarray}
  S_{tot}=\int d^dx \left[-\frac{1}{4}F^{ab} (\partial^l\partial^l)^k  F^{ab}-b (\partial^l\partial^l)^k\partial^a\phi^a
  +\frac{1}{2}b  (\partial^l\partial^l)^k b +\bar c(\partial^l\partial^l)^{k+1} c \right],\ \ 
   k\equiv\frac{d-4}{2},\label{stot}
  \end{eqnarray}
  where field-strength  $F^{ab}=\partial^a\phi^b -\partial^b\phi^a$. Here $\phi^a, b, c$ and $\bar c$
  are spin-1 conformal field, Nakanishi-Lautrup field, ghost field and antighost field respectively.
  In terms of gauge-fixing fermion
  the above action can be described by
  \begin{eqnarray}
  S_{tot}=\int d^dx  \left[-\frac{1}{4}F^{ab} (\partial^l\partial^l)^k  F^{ab} +s_b \Psi^L\right], 
  \end{eqnarray}
  where $\Psi^L=\bar c\left[-(\partial^l\partial^l)^k\partial^a\phi^a+\frac{1}{2}   (\partial^l\partial^l)^k b\right]$.
  The fermionic rigid BRST transformations of the  fields are
  \begin{eqnarray}
  s_b\phi^a =-\partial^a c, \ \ s_b b=0,\ \ s_b c=0,\ \ s_b\bar c=b.\label{brs}
  \end{eqnarray}
 The generating functional for spin-1 conformal field theory corresponding to (\ref{stot})
is defined by
\begin{eqnarray}
Z^L[0] =\int {\cal D}\phi^a{\cal D}b{\cal D}c{\cal D}\bar c\exp({iS_{tot}}).
\end{eqnarray}

However, the BRST invariant action for spin-1 conformal field in non-linear (quadratic) gauge is given by
  \begin{eqnarray}
  S_{tot}^{quad}&=&\int d^dx \left[-\frac{1}{4}F^{ab} (\partial^l\partial^l)^k  F^{ab}-b (\partial^l\partial^l)^k\partial^a\phi^a-b (\partial^l\partial^l)^k\phi^a\phi^a
  +\frac{1}{2}b  (\partial^l\partial^l)^k b 
  \right. \nonumber\\
  &+&\bar c(\partial^l\partial^l)^{k+1} c+ \left. 2\bar c(\partial^l\partial^l)^k\phi^a\partial^a c\right],\nonumber\\
  &=&\int d^dx \left[-\frac{1}{4}F^{ab} (\partial^l\partial^l)^k  F^{ab}+s_b\Psi^{NL}\right], 
  \nonumber\\
  &=&\int d^dx \left[-\frac{1}{4}F^{ab} (\partial^l\partial^l)^k  F^{ab}+s_b\left(\bar c\left[-(\partial^l\partial^l)^k\partial^a\phi^a -(\partial^l\partial^l)^k\phi^a\phi^a +\frac{1}{2}   (\partial^l\partial^l)^k b\right]\right)\right],\end{eqnarray}
  which remains invariant under same set of BRST transformations given in (\ref{brs}).
Following the method given in section II, we construct the FFBRST transformation as follows: 
    \begin{eqnarray}
  \delta_f\phi^a =-\partial^a c\ \Theta[\varphi^{a_1}], \ \ \delta_f b=0,\ \ \delta_f c=0,\ \ \delta_f\bar c=b\ \Theta[\varphi^{a_1}],\label{brst}
  \end{eqnarray}
  where $\Theta[\varphi^{a_1}]$ is an arbitrary finite field-dependent BRST parameter.
  
  Now, we construct a particular $\Theta[\varphi^{a_1}]$ to calculate the Jacobian for path integral measure whose infinitesimal version  is evaluated as follows
  \begin{eqnarray}
\Theta'[\varphi^{a_1}] =-i\int d^dx\left[\bar c(\partial^l\partial^l)^k \phi^a\phi^a  \right].
\end{eqnarray}
Now we calculate the change in Jacobian with respect to continuous parameter $\kappa$ as follows
\begin{eqnarray}
\frac{1}{J(\kappa)}\frac{dJ(\kappa)}{d\kappa} = i\int d^d x\left[ -b(\partial^l\partial^l)^k \phi^a\phi^a
+2\bar c (\partial^l\partial^l)^k\phi^a\partial^ac \right],\label{jaco}
\end{eqnarray}
where we have utilized the relation (\ref{jac}).

To exponentiate the Jacobian we propose the following local functional
\begin{eqnarray}
S_1[\varphi^{a_1}] = \int d^dx\left[\xi_1 b(\partial^l\partial^l)^k \phi^a\phi^a
+\xi_2\bar c (\partial^l\partial^l)^k\phi^a\partial^ac \right],\label{s1}
\end{eqnarray}
where $\xi_1$ and $\xi_2$ are $\kappa$-dependent arbitrary constant parameters.
The equations (\ref{jaco}) and (\ref{s1}) together with (\ref{mcond}) yields
the following linear differential equations: 
\begin{eqnarray}
\xi_1^\prime +1=0,\ \ \xi_2^\prime -2=0.
\end{eqnarray}
The exact solutions of the above equations satisfying the boundary condition ($\xi_i(\kappa=0)=0$) are
\begin{eqnarray}
\xi_1  =-\kappa,\ \ \xi_2 = 2\kappa.
\end{eqnarray}
With these identifications the expression of local functional becomes 
\begin{eqnarray}
S_1[\varphi^{a_1}]  =  \int d^dx\left[ -\kappa b(\partial^l\partial^l)^k \phi^a\phi^a
+2\kappa\bar c (\partial^l\partial^l)^k\phi^a\partial^ac \right].
\end{eqnarray}
This is evident from above expression that at $\kappa=0$ the functional $S_1$ vanishes.
However, at $\kappa=1$ this takes the following form:
\begin{eqnarray}
S_1[\varphi^{a_1}]_{\kappa=1}  =  \int d^dx\left[ -b(\partial^l\partial^l)^k \phi^a\phi^a
+2\bar c (\partial^l\partial^l)^k\phi^a\partial^ac \right].
\end{eqnarray}
So, according to (\ref{ext}), after performing the FFBRST transformation on generating functional 
the effective action (\ref{stot}) modifies by
\begin{eqnarray}
S_{tot}+S_1[\varphi^{a_1}]_{\kappa=1} =S_{tot}^{quad}.
\end{eqnarray}
Therefore, we observe that the FFBRST transformation on generating functional of spin-1 conformal
theory in linear gauge
changes the effective action from linear gauge to quadratic gauge within functional integral.
Here we note that the FFBRST transformation amounts the precise change on the BRST exact part of
the effective action. We construct the finite parameter in such a manner that Jacobian 
of the path integral measure amounts change in the BRST-exact part of the effective action.
  \subsection{Spin-2 conformal field}
The classical action for spin-2 conformal field theory  (a particular form of (\ref{tot})) is given by 
  \begin{eqnarray}
   S_{inv}=\int d^dx \left [R^{ab}_{lin}(\partial^l\partial^l)^{k-1}R^{ab}_{lin} -\frac{d}{4(d-1)}
    R_{lin}(\partial^l\partial^l)^{k-1}R_{lin}\right],\ \ k\equiv\frac{d-2}{2},
   \end{eqnarray}
  where $R^{ab}$ is expressed by
     \begin{eqnarray}
    R^{ab}_{lin} =\frac{1}{2}\left(-(\partial^l\partial^l)\phi^{ab} +\partial^a\partial^c\phi^{cb}
   +\partial^b\partial^c\phi^{ca}-\partial^a\partial^b\phi^{cc}\right).
   \end{eqnarray} 
   The gauge-fixing and ghost action is given together by 
    \begin{eqnarray}
   S_{gf}&=&\int d^dx \left[ -b^a(\partial^l\partial^l)^k(\partial^b\phi^{ab}-\frac{1}{2}\partial^a\phi^{bb})+\frac{1}{u^2}(b-\partial^a b^a)(\partial^l\partial^l)^{k-1}(\partial^c\partial^e\phi^{ce} -(\partial^l\partial^l)\phi^{cc})\right.\nonumber\\
   &+& \left. b^a(\partial^l\partial^l)^k b^a+\frac{1}{2u^2}(b-\partial^a b^a)(\partial^l\partial^l)^{k-1}(b-\partial^c b^c)+ \bar c^a(\partial^l\partial^l)^{k+1} c^a +\bar c (\partial^l\partial^l)^k c\right].
    \end{eqnarray}
    So, the complete action is given by
    \begin{eqnarray}
    S_{tot} =S_{inv} +S_{gf},
    \end{eqnarray}
    which is invariant under following  BRST transformation:
 \begin{eqnarray}
 \delta_b \phi^{ab}&=&-\left(\partial^ac^b +\partial^b c^a +\frac{2}{d-2}\eta^{ab}c\right)\delta\lambda,\nonumber\\
 \delta_b\phi^a &=&-(\partial^a c-\partial^l\partial^l c^a)\delta\lambda,\nonumber\\
 \delta_b\phi  &=& u\partial^l\partial^l c\ \delta\lambda,\nonumber\\
  \delta_b  c^a &=&0,\ \  \delta_b c=0,\nonumber\\
    \delta_b\bar c^a &=&b^a\ \delta\lambda,\ \  \delta_b c=b\ \delta\lambda,\nonumber\\
      \delta_b b^a &=&0,\ \  \delta_b b=0,
 \end{eqnarray}
 where $\delta\lambda$ infinitesimal, anticommuting parameter.
 The FFBRST transformation is constructed by
  \begin{eqnarray}
 \delta_f \phi^{ab}&=&-\left(\partial^ac^b +\partial^b c^a +\frac{2}{d-2}\eta^{ab}c\right)\Theta[\varphi^{a_1a_2}],\nonumber\\
 \delta_f\phi^a &=&-(\partial^a c-\partial^l\partial^l c^a)\Theta[\varphi^{a_1a_2}],\nonumber\\
 \delta_f\phi  &=& u\partial^l\partial^l c\ \Theta[\varphi^{a_1a_2}],\nonumber\\
  \delta_f  c^a &=&0,\ \  \delta_b c=0,\nonumber\\
    \delta_f\bar c^a &=&b^a\ \Theta[\varphi^{a_1a_2}],\ \  \delta_f c=b\ \Theta[\varphi^{a_1a_2}],\nonumber\\
      \delta_f b^a &=&0,\ \  \delta_f b=0.
 \end{eqnarray}
 To construct the finite field-dependent parameter $\Theta[\varphi^{a_1a_2}]$ we choose the following 
 infinitesimal parameter:
 \begin{eqnarray}
\Theta'[\varphi^{a_1a_2}] =-i\int d^dx\left[ \bar c^a (\partial^l\partial^l)^k\left(\phi^b\phi^{ab}-\frac{1}{2}\phi^a
\phi^{bb}\right)\right].
\end{eqnarray}
The change in Jacobian under FFBRST transformation is calculated by
\begin{eqnarray}
\frac{1}{J(\kappa)}\frac{dJ(\kappa)}{d\kappa}& =& i\int d^dx\left[ -b^a(\partial^l\partial^l)^k \left(\phi^b\phi^{ab} -\frac{1}{2}\phi^a\phi^{bb}\right) 
+ \bar c^a (\partial^l\partial^l)^k \partial^bc\phi^{ab} \right.\nonumber\\
&-&\left.\bar c^a (\partial^l\partial^l)^{k+1}c^b\phi^{ab}
+\bar c^a (\partial^l\partial^l)^k\partial^a c^b\phi^b +\bar c^a (\partial^l\partial^l)^k\partial^b c^a\phi^b \right.\nonumber\\
&-&\left. \frac{2}{d-2}\bar c^a (\partial^l\partial^l)^k\eta^{ab}\phi^b
-\frac{1}{2}\bar c^a (\partial^l\partial^l)^k\partial^a c\phi^{bb}
+ 
\frac{1}{2}\bar c^a  (\partial^l\partial^l)^{k+1}c^a\phi^{bb}
\right.\nonumber\\
&-&\left.\bar c^a (\partial^l\partial^l)^k \partial^bc^b\phi^a -\frac{1}{d-2}\bar c^ (\partial^l\partial^l)^k
\eta^{bb} c\phi^a\right].\label{jaco1}
\end{eqnarray}  
Keeping the forms of effective action in linear and quadratic gauges in mind we make an ansatz for $S_1$ in this case as follows
\begin{eqnarray}
S_1 [\varphi^{a_1a_2}] & =&  \int d^dx\left[ \xi_1 (\kappa) b^a(\partial^l\partial^l)^k \left(\phi^b\phi^{ab} -\frac{1}{2}\phi^a\phi^{bb}\right) 
+ \xi_2 (\kappa)  \bar c^a (\partial^l\partial^l)^k \partial^bc\phi^{ab} \right.\nonumber\\
&+&\left. \xi_3 (\kappa) \bar c^a (\partial^l\partial^l)^{k+1}c^b\phi^{ab}
+  \xi_4 (\kappa) \bar c^a (\partial^l\partial^l)^k\partial^a c^b\phi^b + \xi_5 (\kappa) \bar c^a (\partial^l\partial^l)^k\partial^b c^a\phi^b \right.\nonumber\\
&+&\left. \xi_6 (\kappa) \bar c^a (\partial^l\partial^l)^k\eta^{ab}\phi^b
+ \xi_7 (\kappa) \bar c^a (\partial^l\partial^l)^k\partial^a c\phi^{bb}
+ 
 \xi_8 (\kappa) \bar c^a  (\partial^l\partial^l)^{k+1}c^a\phi^{bb}
\right.\nonumber\\
&+&\left. \xi_9 (\kappa) \bar c^a (\partial^l\partial^l)^k \partial^bc^b\phi^a +
 \xi_{10} (\kappa) \bar c^ (\partial^l\partial^l)^k
\eta^{bb} c\phi^a\right].\label{s2}
\end{eqnarray}
The essential condition (\ref{mcond}) together with (\ref{jaco1}) and (\ref{s2}) yields
the following differential equations for $\xi_i$:
\begin{eqnarray}
&&\xi_1' +1=0,\ \ \xi_2'-1=0,\ \ \xi_3' +1=0,\ \ \xi_4' -1=0,\nonumber\\
&&\xi_5' -1=0,\ \ \xi_6'+\frac{2}{d-2}=0,\ \ \xi_7' +\frac{1}{2}=0,\ \ \xi_8' -\frac{1}{2}=0,\nonumber\\
 &&\xi_9' +1=0,\ \ \xi_{10}'+\frac{1}{d-2}=0.
\end{eqnarray}
The exact solutions of these differential equations satisfying boundary condition ($\xi_i(\kappa=0)=0$) are given by
\begin{eqnarray}
&&\xi_1  =-\kappa,\ \ \xi_2  =\kappa,\ \ \xi_3' =-\kappa,\ \ \xi_4'  =\kappa,\nonumber\\
&&\xi_5'  =\kappa,\ \ \xi_6'=-\frac{2}{d-2}\kappa,\ \ \xi_7' =-\frac{1}{2}\kappa,\ \ \xi_8' =\frac{1}{2}\kappa,\nonumber\\
 &&\xi_9' =-\kappa,\ \ \xi_{10}'=-\frac{1}{d-2}\kappa.
\end{eqnarray}
With this solutions the expression of $S_1$ reduces to
\begin{eqnarray}
S_1[\varphi^{a_1a_2}]& =&  \int d^dx\left[ -\kappa b^a(\partial^l\partial^l)^k \left(\phi^b\phi^{ab} -\frac{1}{2}\phi^a\phi^{bb}\right) 
+\kappa \bar c^a (\partial^l\partial^l)^k \partial^bc\phi^{ab} \right.\nonumber\\
&-&\left.\kappa\bar c^a (\partial^l\partial^l)^{k+1}c^b\phi^{ab}
+ \kappa\bar c^a (\partial^l\partial^l)^k\partial^a c^b\phi^b +\kappa\bar c^a (\partial^l\partial^l)^k\partial^b c^a\phi^b\right.\nonumber\\
&-&\left.\frac{2}{d-2}\kappa\bar c^a (\partial^l\partial^l)^k\eta^{ab}\phi^b
-\frac{1}{2}\kappa\bar c^a (\partial^l\partial^l)^k\partial^a c\phi^{bb}
+\frac{1}{2}\kappa\bar c^a  (\partial^l\partial^l)^{k+1}c^a\phi^{bb}
\right.\nonumber\\
&-&\left. \kappa\bar c^a (\partial^l\partial^l)^k \partial^bc^b\phi^a -\frac{1}{d-2}\kappa\bar c^ (\partial^l\partial^l)^k
\eta^{bb} c\phi^a\right],
\end{eqnarray}
which vanishes for $\kappa=0$. However, for $\kappa=1$ this  reduces to
\begin{eqnarray}
S_1[\varphi^{a_1a_2}]_{\kappa=1}& =&  \int d^dx\left[ -b^a(\partial^l\partial^l)^k \left(\phi^b\phi^{ab} -\frac{1}{2}\phi^a\phi^{bb}\right) 
+ \bar c^a (\partial^l\partial^l)^k \partial^bc\phi^{ab} \right.\nonumber\\
&-&\left.\bar c^a (\partial^l\partial^l)^{k+1}c^b\phi^{ab}
+ \bar c^a (\partial^l\partial^l)^k\partial^a c^b\phi^b +\bar c^a (\partial^l\partial^l)^k\partial^b c^a\phi^b\right.\nonumber\\
&-&\left.\frac{2}{d-2}\bar c^a (\partial^l\partial^l)^k\eta^{ab}\phi^b
-\frac{1}{2}\bar c^a (\partial^l\partial^l)^k\partial^a c\phi^{bb}
+\frac{1}{2}\bar c^a  (\partial^l\partial^l)^{k+1}c^a\phi^{bb}\right.\nonumber\\
&-&\left. \bar c^a (\partial^l\partial^l)^k \partial^bc^b\phi^a -\frac{1}{d-2}\bar c^ (\partial^l\partial^l)^k
\eta^{bb} c\phi^a\right].
\end{eqnarray} 
Now, after performing the FFBRST transformation  the extended action for spin-2 conformal field, as mentioned in (\ref{ext}),
is calculated by
\begin{eqnarray}
S_{tot}+S_1[\varphi^{a_1a_2}]_{\kappa=1} &=&\int d^dx \left [R^{ab}_{lin}(\partial^l\partial^l)^{k-1}R^{ab}_{lin} -\frac{d}{4(d-1)}
    R_{lin}(\partial^l\partial^l)^{k-1}R_{lin}
    \right.\nonumber\\
   &-& \left.b^a(\partial^l\partial^l)^k(\partial^b\phi^{ab}-\frac{1}{2}\partial^a\phi^{bb})+\frac{1}{u^2}(b-\partial^a b^a)(\partial^l\partial^l)^{k-1}(\partial^c\partial^e\phi^{ce} \right.\nonumber\\
   &-& (\partial^l\partial^l)\phi^{cc})+\left. b^a(\partial^l\partial^l)^k b^a+\frac{1}{2u^2}(b-\partial^a b^a)(\partial^l\partial^l)^{k-1}(b-\partial^c b^c)\right.\nonumber\\
&+&\left.  \bar c^a(\partial^l\partial^l)^{k+1} c^a +\bar c (\partial^l\partial^l)^k c - b^a(\partial^l\partial^l)^k \left(\phi^b\phi^{ab} -\frac{1}{2}\phi^a\phi^{bb}\right) 
   \right.\nonumber\\
   &+ & \left.
\bar c^a (\partial^l\partial^l)^k \partial^bc\phi^{ab} -\bar c^a (\partial^l\partial^l)^{k+1}c^b\phi^{ab}
+\bar c^a (\partial^l\partial^l)^k\partial^a c^b\phi^b \right.\nonumber\\
&+&\left. \bar c^a (\partial^l\partial^l)^k\partial^b c^a\phi^b-\frac{2}{d-2}\bar c^a (\partial^l\partial^l)^k\eta^{ab}\phi^b
-\frac{1}{2}\bar c^a (\partial^l\partial^l)^k\partial^a c\phi^{bb}
\right.\nonumber\\
&+&\left. 
\frac{1}{2}\bar c^a  (\partial^l\partial^l)^{k+1}c^a\phi^{bb}-\bar c^a (\partial^l\partial^l)^k \partial^bc^b\phi^a -\frac{1}{d-2}\bar c^ (\partial^l\partial^l)^k
\eta^{bb} c\phi^a   \right].\label{fi}
\end{eqnarray}
Here we observe that the final action obtained in (\ref{fi}) has non-linear gauge.
Therefore, we observed that the FFBRST transformation relates the generating functionals 
corresponding to linear 
and non-linear gauges for spin-2 conformal field also.
\section{ Conformal field theory in BV formulation}
In this section, we extend the formulation using BV technique.
For this purpose, we need to introduce the antifields ($\varphi^{a_1a_2.....a_{s'}\star}$) corresponding to fields  having opposite statistics
in the configuration space. With the introduction of such antifields, the arbitrary extended quantum action, $W_{\Psi  }[\varphi^{a_1a_2.....a_{s'}},\varphi^{a_1a_2.....a_{s'}\star}] $, satisfies a certain rich mathematical
relation, the so-called  quantum master equation \cite{wein},  which is given by
\begin{equation}
\Delta e^{iW_{\Psi }[\varphi^{a_1a_2.....a_{s'}},\varphi^{a_1a_2.....a_{s'}\star}] } =0,\ \
 \Delta\equiv (-1)^{\epsilon_A
 }\frac{\partial_l}{
\partial\varphi^{a_1a_2.....a_{s'}}}\frac{\partial_l}{\partial\varphi^{a_1a_2.....a_{s'}\star}} ,
\label{mq}
\end{equation}
where $A\equiv(a_1a_2.....a_{s'})$.
Therefore, the extended quantum action $W_{\Psi }$ with different gauge-fixing fermion
$\Psi$ are solutions of the quantum master equation.
We would like to show that FFBRST transformation with appropriate choice
of finite field-dependent parameter relates different 
solutions of quantum master equation.
\subsection{Spin-1 conformal field}
In  terms of field and antifields, the generating functional  for the  spin-1 conformal field theory  
in linear gauge  is defined  by  
\begin{eqnarray}
Z^L[0]  = \int {\cal D}\phi^a{\cal D}b{\cal D}c{\cal D}\bar c\ e^{ i\int d^dx \left[-\frac{1}{4}F^{ab} (\partial^l\partial^l)^k  F^{ab} - \phi_L^{a\star}\partial^a c +\bar c_L^\star b \right]},\label{exl}
\end{eqnarray}
where $ \phi_L^{a\star}$ and $\bar c_L^\star$ are antifields  corresponding to the 
$\phi^a$ and $\bar c$ fields respectively with opposite statistics.
The above generating functional can further be recast compactly as
 \begin{equation}
Z^L [0]= \int {\cal D}\varphi^{a_1}\  e^{ i    W_{\Psi^L  }[\varphi^{a_1},\varphi^{a_1\star}_L] },\label{lan}
\end{equation} 
where $ W_{\Psi^L}[\varphi^{a_1},\varphi^{a_1\star}_L] $ is an extended quantum action (a solution of the quantum master equation defined later) for the conformal theory in linear gauge
and   $\varphi^{a_1\star}_L$ refers to the antifields generically corresponding to the collective field  $\varphi^{a_1}( \equiv 
\phi^a, b, \bar c, c)$.

It is well-known that the antifields    for a gauge theory can explicitly be computed from the   gauge-fixed fermion.  For the conformal  theory in linear  gauge the antifields are 
computed  for the gauge-fixed fermion $\Psi^L=\bar c\left[-(\partial^l\partial^l)^k\partial^a\phi^a  +\frac{1}{2}   (\partial^l\partial^l)^k b\right]$ as following:
 \begin{eqnarray}
\phi_L^{a\star }&=&\frac{\delta\Psi^L }{\delta \phi^{a}}=(\partial^l\partial^l)^k\partial^a\bar c,
\nonumber\\
b_L^{\star}&=&\frac{\delta\Psi^L}{\delta b}= \frac{1}{2}   (\partial^l\partial^l)^k \bar c,\nonumber\\ 
 \bar{c}_L^{\star}&=&\frac{\delta\Psi^L}{\delta \bar{c}}=  -(\partial^l\partial^l)^k\partial^a\phi^a  +\frac{1}{2}   (\partial^l\partial^l)^k b,\nonumber\\ 
 c_L^{\star}&=&\frac{\delta\Psi^L}{\delta c}=0.
\end{eqnarray}
With these identifications of antifields the extended quantum action in (\ref{exl}) coincides with
the total effective action (\ref{stot}).
However, for the non-linear gauge the gauge-fixing fermion   is given by
\begin{equation}
\Psi^{NL}=\bar c\left[-(\partial^l\partial^l)^k\partial^a\phi^a -(\partial^l\partial^l)^k\phi^a\phi^a +\frac{1}{2}   (\partial^l\partial^l)^k b\right].
\end{equation} 
The antifields for the above gauge-fixing fermion  are  estimated by:
 \begin{eqnarray}
\phi_{NL}^{a\star }&=&\frac{\delta\Psi^{NL} }{\delta \phi^{a}}=(\partial^l\partial^l)^k\partial^a\bar c-2(\partial^l\partial^l)^k\phi^a\bar c,
\nonumber\\
b_{NL}^{\star}&=&\frac{\delta\Psi^{NL}}{\delta b}=   \frac{1}{2}   (\partial^l\partial^l)^k \bar c,\nonumber\\ 
 \bar{c}_{NL}^{\star}&=&\frac{\delta\Psi^{NL}}{\delta \bar{c}}=  -(\partial^l\partial^l)^k\partial^a\phi^a -(\partial^l\partial^l)^k\phi^a\phi^a +\frac{1}{2}   (\partial^l\partial^l)^k b,\nonumber\\ 
 c_{NL}^{\star}&=&\frac{\delta\Psi^{NL}}{\delta c}=0.
\end{eqnarray}
Likewise the linear gauge case, the  generating functional for the spin-1 conformal theory in non-linear gauge 
can   be written in compact form as
 \begin{equation}
Z^{NL}[0] = \int {\cal D}\varphi^{a_1}\  e^{ i    W_{\Psi^{NL}  }[\varphi^{a_1},\varphi^{a_1\star}_{NL}] },
\end{equation} 
where $ W_{\Psi^{NL}  }[\varphi^{a_1},\varphi^{a_1\star}_{NL}] $ is an extended quantum action (another solution of the quantum master equation) corresponding to  non-linear gauge.

Now we construct the infinitesimal field/antifield dependent parameter as follows
\footnote{We note in this case that the antifields depend on fields as these expressed in terms
of gauge-fixing fermion. Therefore
this field/antifield dependent parameter actually depends on field only  \cite{alex}.}
\begin{eqnarray}
\Theta'[\varphi^{a_1},\varphi^{a_1\star}]=-i\int d^dx\  \bar c \left(  \bar{c}_{ L}^{\star} - \bar{c}_{NL}^{\star}\right).
\end{eqnarray}
From this infinitesimal parameter the finite field/antifield dependent parameter
can be calculated using the relation (\ref{80}). The FFBRST transformation with such field/antifield
dependent parameter leads to the following Jacobian in the integrand of  functional integral
\begin{eqnarray}
J[\varphi^{a_1a_2.....a_{s'}}(x)] =e^{i \int d^dx\left[  - \phi_{NL}^{a\star}\partial^a c +\bar c_{NL}^\star b  +\phi_L^{a\star}\partial^a c -\bar c_L^\star b  \right]},
\end{eqnarray}
which switches the generating functional of spin-1 conformal theory from one gauge to anther.

Therefore, we establish the  connection of the different solutions ($W_{\Psi^L}$ and $W_{\Psi^{NL}}$) of the quantum master equation at quantum level through FFBRST transformation with appropriately 
constructed finite field-dependent parameter.
\subsection{Spin-2 conformal field}
Introducing the antifields corresponding to fields, the generating functional  for the  spin-2 conformal field theory  
in linear gauge  is defined  by  
\begin{eqnarray}
Z^L[0]  &= &\int {\cal D}\phi^{ab}{\cal D}\phi^a{\cal D}\phi {\cal D}b^a{\cal D}b{\cal D}c^a{\cal D}c{\cal D}\bar c^a{\cal D}\bar c\ \exp\left[ i\int d^dx \left(R^{ab}_{lin}(\partial^l\partial^l)^{k-1}R^{ab}_{lin} \right.\right.\nonumber\\
    &-&\left.\left. \frac{d}{4(d-1)}
    R_{lin}(\partial^l\partial^l)^{k-1}R_{lin}+\varphi^{a_1a_2\star}_L (s_b\varphi^{a_1a_2})\right) 
    \right],\label{exl1}
\end{eqnarray}
where $\varphi^{a_1a_2\star}$ are antifields  corresponding to the 
$\varphi^{a_1a_2}( \equiv 
\phi^{ab}, \phi^a, \phi, b^a, b, c^a, c, \bar c^a, \bar c)$ fields generically with opposite statistics.
This can further be written in compact notation as
 \begin{equation}
Z^L[0] = \int {\cal D}\varphi^{a_1a_2}\  e^{ i    W_{\Psi^L  }[\varphi^{a_1a_2},\varphi^{a_1a_2 \star}_L] },\label{lan}
\end{equation} 
where $W_{\Psi^L  }[\varphi^{a_1a_2},\varphi^{a_1a_2 \star}_L] $ is the extended quantum action for spin-2
conformal theory in linear gauge.

In the same fashion, we define the  generating functional for the spin-2 conformal theory for non-linear gauge   in compact form as
 \begin{equation}
Z^{NL}[0]  = \int {\cal D}\varphi^{a_1a_2}\  e^{ i    W_{\Psi^{NL}  }[\varphi^{a_1a_2},\varphi^{a_1a_2 \star}_{NL}] },
\end{equation} 
where $ W_{\Psi^{NL}  }[\varphi^{a_1a_2},\varphi^{a_1a_2 \star}_{NL}]$ is the extended quantum action  corresponding to  non-linear gauge.

We construct the infinitesimal field/antifield dependent parameter for this case as follows:
\begin{eqnarray}
\Theta'[\varphi^{a_1a_2},\varphi^{a_1a_2 \star}]=-i\int d^dx\ [\bar c^a \left(  \bar{c}_{ L}^{a\star} - \bar{c}_{NL}^{a\star}   \right)+ \bar c \left(  \bar{c}_{ L}^{\star} - \bar{c}_{NL}^{\star}   \right)].
\end{eqnarray}
The finite field/antifield dependent parameter
can be evaluated from relation (\ref{80}). The FFBRST transformation with such field/antifield
dependent parameter leads to the following Jacobian in the integrand of  functional integral
\begin{eqnarray}
J[\varphi^{a_1a_2.....a_{s'}}(x)] =e^{i \int d^dx\left[ \varphi^{a_1a_2\star}_{NL} (s_b\varphi^{a_1a_2}) -\varphi^{a_1a_2\star}_L (s_b\varphi^{a_1a_2}) \right]},
\end{eqnarray}
which transforms the generating functional of spin-2 conformal theory from linear gauge to non-linear.
Hence,  the  connection of the different solutions ($W_{\Psi^L}$ and $W_{\Psi^{NL}}$) of the quantum master equation   for spin-2  is established through FFBRST transformation 
with properly constructed parameter. In fact any two solutions
of quantum master equation are connected through 
FFBRST transformation with different finite parameter.
\section{Conclusions}
In this paper we have developed the FFBRST transformation for arbitrary spin-s conformal field theory.
We construct the FFBRST transformation by making the transformation parameter 
finite and field-dependent.  The parameter is made finite and field-dependent by  making all the fields
first (a continuous constant parameter) $\kappa$-dependent and then define a infinitesimal field-dependent
BRST transformation. After that we integrate the parameter of infinitesimal field-dependent BRST transformation in the limiting values of $\kappa$ which yields the finite field-dependent BRST parameter.
   The novelty of the FFBRST transformation is that 
it leads to a local Jacobian for path integral measure and this Jacobian amounts 
a change in the BRST exact part of the effective action.
Here we note that analogous to  ordinary (non-conformal) quantum field theories 
 the resulting Jacobian in the case of conformal field theories  are still
 local  in nature. This assures the consistancy  of generalized BRST formulation   for
 CFTs also. 
For illustration purpose, we have considered the spin-1 and spin-2 conformal theories.  
For such theories we have
explicitly constructed the specific finite field-dependent parameters.
Furthermore, we have found that the Jacobians corresponding to such parameters switches the theories
from one  gauge to another (namely, linear to non-linear gauges).
Furthermore, we have established the theory at quantum level by analysing it through BV formulation.
In  BV formulation we have demonstrated that the finite field dependent BRST transformation
connects the different solutions of quantum master equation for both spin-1 and spin-2 conformal theory.
Thus our formulation will be helpful in estimating the observables of the conformal theory 
in different gauges.

\end{document}